\documentclass[a4paper,fleqn,usenatbib]{mnras}

\usepackage[T1]{fontenc}
\usepackage{ae,aecompl}

\usepackage{graphicx}	
\usepackage{amsmath}	
\usepackage{amssymb}	

\title[Decretion discs in Be/X-ray binaries]{Modelling decretion discs in Be/X-ray binaries}

\author[R. O. Brown et al.]{
R. O. Brown,$^{1}$\thanks{E-mail: rob1g10@soton.ac.uk}
M. J. Coe,$^{1}$
W. C. G. Ho,$^{1, 2, 3}$
A. T. Okazaki$^{4}$
\\
$^{1}$Physics and Astronomy, University of Southampton, Southampton, SO17 1BJ, UK \\
$^{2}$Mathematical Sciences and STAG Research Centre, University of Southampton, Southampton, SO17 1BJ, UK  \\
$^{3}$Department of Physics and Astronomy, Haverford College, 370 Lancaster Avenue, Haverford, PA 19041, USA \\
$^{4}$Faculty of Engineering, Hokkai-Gakuen University, Toyohira-ku, Sapporo, 062-8605, Japan
}

\date{Accepted XXX. Received YYY; in original form ZZZ}

\pubyear{2018}

\begin{document}
\label{firstpage}
\pagerange{\pageref{firstpage}--\pageref{lastpage}}
\maketitle

\begin{abstract}

As the largest population of high mass X-ray binaries, Be/X-ray binaries provide an excellent laboratory to investigate the extreme physics of neutron stars. It is generally accepted that Be stars possess a circumstellar disc, providing an additional source of accretion to the stellar winds present around young hot stars. Interaction between the neutron star and the disc is often the dominant accretion mechanism. A large amount of work has gone into modelling the properties of these circumstellar discs, allowing for the explanation of a number of observable phenomena. In this paper, smoothed particle hydroynamics simulations are performed whilst varying the model parameters (orbital period, eccentricity, the mass ejection rate of the Be star and the viscosity and orientation of the disc). The relationships between the model parameters and the disc's characteristics (base gas density, the accretion rate of the neutron star and the disc's size) are presented. The observational evidence for a dependency of the size of the Be star's circumstellar disc on the orbital period (and semi-major axis) is supported by the simulations. 

\end{abstract}




\section{Introduction}

Be stars are B spectral type stars that are non-supergiant and have, or have had at some time, one or more Balmer lines in emission \citep{JascEgre1982}. Relative to many other main sequence stars, Be stars have high rates of mass outflow, and are often observed to possess high rotational velocities \citep{Slettebak1988}. The combination of non-radial pulsations and rapid rotation is thought to cause material to be ejected from the surface of the Be star, leading to a diffuse and gaseous circumstellar disc \citep{Riv2000}. This is commonly referred to as a decretion disc. Given that stellar winds from Be stars are often weak, the cirumstellar disc provides the dominant source of material for accretion onto a binary partner. If the Be star is accompanied by a compact object, the system is a High Mass X-ray Binary (HMXB). An upper limit to the decretion rate of particular Be stars was initially determined to be $\dot{M} = 10^{-8} $M$_{\odot}$yr$^{-1}$ \citep{Riv2000} but has more recently been refined to the range of $10^{-12} - 10^{-9}$ M$_{\odot}$yr$^{-1}$ \citep{Vieira2015}. \citet{Ghor2018} suggest that the mass ejection rate of $\omega$ CMa reaches $\sim4 \times 10^{-7}$M$_{\odot}$yr$^{-1}$ by modelling the observed V-band light curve. It should be noted however, that $\omega$ CMa is one of the brightest Be stars that can be observed ($m_{\text{V}} \approx 3.6$ to $4.2$). The typical mass decretion rate of the central star for observable (and likely more dense) Be star discs has been shown to be $\dot{M} = 10^{-10}$ M$_{\odot}$yr$^{-1}$ by \citet{Rimulo2018}. 

Be/X-ray binaries are the largest population of observable HMXBs \citep{RapHeu1982, HeuRap1987, Coleiro2013}. The varying size of the disc, coupled with the interaction of a compact object leads to a variety of observable effects. Some of these, such as superorbital modulation and giant outbursts, are not well understood. Discerning the exact nature of these phenomena can lead to a better understanding of the extreme physics of neutron stars and black holes (see \citet{Reig2011} for review).

\citet{CoeKirk2015} (hereafter referred to as CK15) investigated a sample of approximately 70 X-ray emitting binary systems containing a Be star in the Small Magellanic Cloud (SMC). All these Be/X-ray binaries show clear X-ray pulse signatures from a neutron star. In the paper, they list all the known orbital periods, eccentricities and neutron star spin periods. Also included is the Be star's spectral type, the size of the circumstellar disc and evidence for non-radial pulsations. The relationship between orbital period and circumstellar disc size seen in galactic sources \citep{ReigFabCoe1997} is shown to be clearly present in the SMC systems. Figures 8 and 10 in CK15 show two correlations that will be investigated in Section \ref{sec:Obs}.

\citet{Panoglou2016} examined the size, shape and density structure of coplanar decretion discs in simulations targeting the Be/X-ray binary system, $\zeta$ Tauri. They use the parameter values from \citet{Ruzdjak2009} but use both the real orbital period of 133 days and an alternate period of 30 days. This is because tidal effects are much greater in close binaries. \citet{Cyr2017} continue this investigation but apply the same methods to misaligned discs in a Be/X-ray binary with a B5 star. 

In this paper, simulations are used to probe fundamental characteristics of Be/neutron star binaries and generally expands upon the nature of the work conducted in \citet{Panoglou2016} and \citet{Cyr2017}. In Section \ref{sec:Simulations}, the methods used to model Be/neutron star binaries are discussed. Sections \ref{sec:BGD}, \ref{sec:discSize} and \ref{sec:NSacc} detail the investigations into the disc's base gas density, the disc's size and the neutron star's accretion rate, respectively. The relationships between the disc's characteristics (the base gas density, the accretion rate of the neutron star companion and the disc's size) and the simulation parameters (the viscosity, the mass ejection rate of the Be star, the orientation of the disc, the orbital period and eccentricity) are presented. In Section \ref{sec:Obs} the simulations are used to test the relationships shown in Figures 8 and 10 of CK15. Section \ref{sec:Conclusions} discusses the results shown in the paper and compares the findings to previous works. 

\section{Simulations} \label{sec:Simulations}

A 3D smoothed particle hydrodynamics code is used to simulate Be/X-ray binaries. \citet{Bate1995} developed the code to model protobinary systems, building upon an already existing smoothed particle hydrodynamics code created by Benz \citep{Benz1990,Benzetal1990}. The code was then modified to simulate viscous decretion discs by \citet{OkazakiTrunc2002}. The Be star's decretion disc is modelled by an ensemble of particles each of mass $\sim10^{-15}$M$_{\odot}$. The disc is assumed to be isothermal with a temperature of $T = 0.6 T_{eff}$ for simplicity \citep{PoeMarl1982, vanKer1995, CarcBjor2006}. The particles are injected into the disc with Keplerian velocity at a random azimuthal angle at 1.05 stellar radii from the centre of the Be star. They are placed at a random small distance from the equatorial plane. The Be star mass and radius are assumed to be 18M$_{\odot}$ and 7R$_{\odot}$, respectively, to target a B0V star \citep{Schmidt1982,Cox2000}. A mass of 1.4M$_{\odot}$ and radius of 10km is adopted for the neutron star. The simulations are evolved until they reach a steady state, i.e. when the number of particles in the disc at apastron changes by less than 1$\%$ for more than 5 orbits. The amount of time the systems take to reach this steady state is dependent on the orbital period and viscosity but these times vary between ten and one hundred orbital periods. The total number of simulation particles in the steady state varies between 10,000 and 100,000 depending on the model parameters. 


There are a number of simulation parameters that are varied individually to explore the aforementioned characteristics of Be/neutron binaries. These include the inclination of the disc to the orbital plane, the mass ejection rate of the Be star, the viscosity of the disc, the binary period and the orbital eccentricity. The behaviour and structure of the disc is dependent on all these parameters. \citet{Rimulo2018} found that the Shakura-Sunyaev viscosity parameter, $\alpha$ \citep{ShakSuny1973}, in the discs of Be stars could be anywhere from a few tenths to more than one. \citet{Ghor2018} similarly used viscosity parameters from $\alpha = 0.1$ to $\alpha = 1$ to model the Be star $\omega$ CMa. Hence, the viscosity explored here is in the range $0.1 \le \alpha \le 1.5$. The mass ejection rate of the Be star is varied from $10^{-11}$ to $10^{-5}$M$_{\odot}$yr$^{-1}$ to encompass the large mass ejection rates used by \citet{Ghor2018} and to include ejection rates inside the range suggested by \citet{Vieira2015}. It should be noted that the mass ejection rate of the Be star is higher than the mass-loss/decretion rate due to the material that is immediately reaccreted onto the Be star \citep{Haubois2012}. The sample of Be/neutron star binaries used by CK15 possess orbital periods of up to $\sim500$ days. The orbital period of the simulations is varied from 40 days to 400 days in steps of 40 days with the aim to broadly cover this range.  Eccentricity is tested at $e=0.0, 0.2, 0.4$ and $0.6$ for the systems of varying orbital period and $e=0.0, 0.2$ and $0.4$ when varying disc orientation (see Section \ref{sec:misalignment}). For the simulations where the viscosity of the disc and the mass ejection rate of the Be star are varied, the eccentricity is only tested at $e=0.0$ and $0.4$. This range is chosen because over $90\%$ of Be/neutron star binaries of known eccentricity have $e \le 0.6$ \citep{BrownBH2018}.

When the viscosity and mass ejection rate are not specified, these two quantities are assumed to be $\alpha = 0.63$ and $\dot{M} = 10^{-10}$ M$_{\odot}$yr$^{-1}$, respectively. These are typical values for observable Be stars as determined through simulations by \citet{Rimulo2018}. When the period of the simulation is unspecified, an orbital period of 40 days is assumed. This is because the majority of the systems in the observational sample contained in CK15 have orbital periods of $P_{\mathrm{orb}} \le 150$ days. Given the masses of the Be star and neutron star, the semi-major axis of these systems is $\sim19$ stellar radii. The Be star's disc lies in the the orbital plane unless it is stated otherwise.


\subsection{Misaligned discs} \label{sec:misalignment}

\begin{figure}
	\centering
	\includegraphics[width=.5\textwidth]{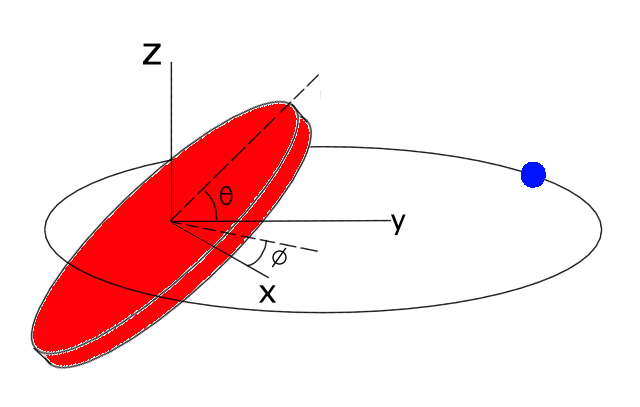}
	\caption{An illustration of the geometry used in this paper. The Be star's circumstellar disc is shown in red and the neutron star is represented by the solid blue circle. The orbit of the neutron star is in the x-y plane. The two angles, $\theta$ and $\phi$, show the inclination and azimuthal angles, respectively. The inclination and azimuthal angle are defined as the rotation about the x axis and z axis, respectively. Periastron and apastron both lie along the y axis.}
	\label{fig:SysGeom}
\end{figure}

There is a subset of the simulations shown in this paper that investigate the effect of changing the orientation of the disc relative to the orbital plane. The disc inclination is investigated because there are a number of Be/X-ray binary systems that have had notable observational features that are thought to arise due to large misalignments between the disc and the orbital plane. Examples of these include GX 304-1 \citep{Postnov2014} and SXP 5.05 \citep{Coe2015,Brown2019}. It is suggested that the Be star's disc also precesses in GX 304-1 \citep{Kuhnel2017} but the investigation of this using the simulations is left for future work. Figure \ref{fig:SysGeom} demonstrates the definition of the two orientation angles in this paper, inclination angle, $\theta$, and azimuthal angle, $\phi$. A disc at an inclination angle of 0$^{\circ}$ lies in the orbital plane and the azimuthal rotation is arbitrary. The x-y plane in the diagram is identical to the orbital plane. 

When the circumstellar disc of a Be star is highly misaligned to the orbital plane, the three-body phenomenon known as the Kozai-Lidov mechanism becomes important \citep{Kozai1962,Lidov1962,MartinBe2014}. This mechanism causes the disc's misalignment to oscillate, trading its inclination for an increase in the eccentricity of the disc. The period of Kozai-Lidov cycles for a particle orbiting the primary of a binary, as derived by \citet{Kiseleva1998}, is given by

\begin{equation} \label{eq:KLperiod}
\frac{\tau_{\mathrm{KL}}}{P_{\mathrm{orb}}} \approx \frac{M_{1} + M_{2}}{M_{2}} \frac{P_{\mathrm{orb}}}{P_{\mathrm{p}}} (1 - e^{2})^{\frac{3}{2}}
\end{equation}

\noindent where $M_{1}$ is the mass of the primary, $M_{2}$ is the mass of the secondary and $P_{\mathrm{p}}$ is the orbital period of the particle. All the simulations in this paper that contain a misaligned disc have the same orbital period and thus have an identical Kozai-Lidov period for the same eccentricity. The time the simulations require to reach a steady state is shorter than the period of the Kozai-Lidov mechanism.

Table \ref{tab:discEcc} shows the average disc eccentricity of twelve simulations with different disc inclinations and orbital eccentricities. The initial inclination angle is the angle between the equatorial plane of the Be star (the same plane that the disc material is ejected into) and the orbital plane. Disc eccentricity shown is averaged over the final five orbits of the simulation. The simulations have been run for similar times and thus, all the systems of equal eccentricity are at the same approximate point in the Kozai-Lidov cycle. Up until the time the simulations have been evolved, the average disc eccentricity is greater for systems with larger initial misalignments between the disc and the orbital plane. This is expected as the systems with a larger initial inclination angle are capable of achieving larger disc eccentricities. When orbital eccentricity is increased, the Kozai-Lidov period is decreased (see Equation \ref{eq:KLperiod}) meaning that the simulation ends at a later stage of the Kozai-Lidov cycle and hence the modelled disc is, on average, more eccentric. The disc is not completely circular even in binaries with a coplanar disc because of the interaction of the neutron star with the disc. The inclination of the disc does not oscillate in coplanar systems. 

\begin{table}
	\centering
	\caption{Disc eccentricities for a number of the simulations with discs misaligned to the orbital plane. Initial inclination angle refers to the angle at which the disc material is injected at, and is not the current inclination of the disc. The definition of the inclination angle is illustrated in Figure \ref{fig:SysGeom}. Disc eccentricity is averaged over the final 5 orbits of the simulation. All of the systems shown in this table have an azimuthal rotation of $\phi = 0^{\circ}$.}
    \label{tab:discEcc}
	\begin{tabular}{ccc} 
		\hline
		orbital $e$ & initial inclination angle ($^{\circ}$) & mean disc $e$ \\
		\hline
		0.0         & 0                        & 0.11             \\
		0.0         & 30                       & 0.15             \\
		0.0         & 60                       & 0.29             \\
		0.0         & 90                       & 0.34             \\
		0.2         & 0                        & 0.15             \\
		0.2         & 30                       & 0.19             \\
		0.2         & 60                       & 0.32             \\
		0.2         & 90                       & 0.43             \\
		0.4         & 0                        & 0.17             \\
		0.4         & 30                       & 0.22             \\
		0.4         & 60                       & 0.36             \\
		0.4         & 90                       & 0.44             \\
		\hline
	\end{tabular}
\end{table}

\section{Base gas density} \label{sec:BGD}

The unperturbed density profile of a Be star's circumstellar disc can be described by a Gaussian-modified power law \citep{TouhGies2011,RivCarc2013}. It has also been shown previously that in a steady state, the decretion disc's surface density is proportional to $\dot{M} / \alpha$ \citep{CarcBjor2008}. The density at the surface of the Be star, i.e. base of the circumstellar decretion disc, is defined as the base gas density. For the simulations discussed in this paper, base gas density is determined by the mean density of an equatorial ring of simulation particles at the Be star's surface, averaged over five orbits. 

\begin{figure}
	\centering
	\includegraphics[width=.5\textwidth]{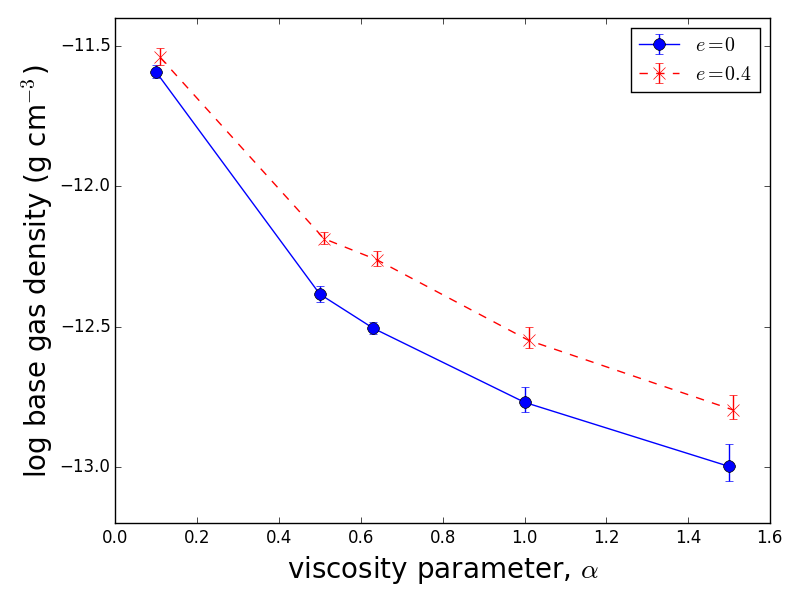}
	\caption{The relationship between the base gas density and the viscosity parameter of the disc. The bars show the minimum and maximum values of the base gas density around an orbital cycle. The bars are comparable to or smaller than the size of the symbols. This is for systems with a 40 day period and eccentricities of $e=0$ and 0.4.}
	\label{fig:VISCbgd}
\end{figure}

Figure \ref{fig:VISCbgd} shows that the base gas density decreases almost exponentially with viscosity. The viscosity of the disc matter leads to the transfer of angular momentum, i.e. matter travels outwards due to the angular momentum gained from the particles falling back onto the star. Hence, as viscous forces are diffusive, a larger viscosity leads to a lower base gas density. The base gas density varies by a factor of $\sim10$ due to viscosity. 

\begin{figure}
	\centering
	\includegraphics[width=.5\textwidth]{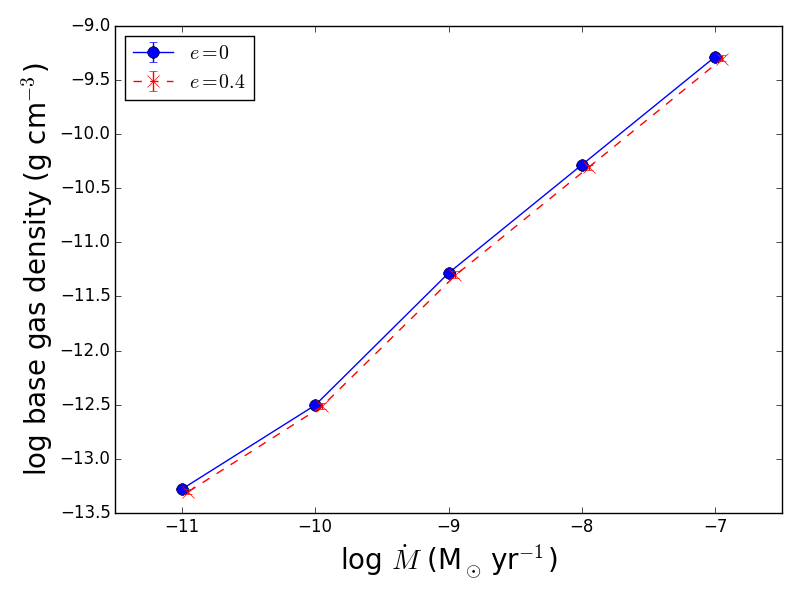}
	\caption{The relationship between the base gas density and the mass ejection rate of the Be star. The bars show the minimum and maximum values of the base gas density around an orbital cycle. The bars are comparable to or smaller than the size of the symbols. This is for systems with a 40 day period and eccentricities of $e=0$ and 0.4.}
	\label{fig:INJEbgd}
\end{figure}

Base gas density increases linearly with the Be star's mass ejection rate as demonstrated by Figure \ref{fig:INJEbgd}. The base gas density at equilibrium arises from the balance between the amount of material being ejected into the disc and the matter falling back onto the Be star. Therefore, the base gas density of the disc is heavily dependent on the mass ejection rate, varying by $\sim4$ orders of magnitude over the range shown.

Base gas density is not related to orbital period, eccentricity or the orientation of the disc. These quantities only alter the neutron star's interaction with the disc and thus do not affect the innermost regions.

\section{Disc size} \label{sec:discSize}

Observations have shown that the circumstellar discs of Be stars can be up to hundreds of stellar radii in size \citep{DouTay1992}. The size of the circumstellar disc in Be/X-ray binaries is dependent on the tidal truncation caused by the binary partner \citep{OkazakiTrunc2002}. In equilibrium, the disc is limited to a radius where the tidal forces balance the viscous forces that is defined as the truncation radius. In this paper, the size of the Be star's circumstellar disc is defined as the radius that contains 90$\%$ of the simulation particles. Note that this definition means that the size of an eccentric disc will be defined by its semi-major axis. Minimum and maximum disc sizes are calculated for individual orbits and then averaged over five orbits. In all plots in this section, solid lines and dashed lines represent the maximum and minimum values of disc size, respectively. 

\begin{figure}
	\centering
	\includegraphics[width=.5\textwidth]{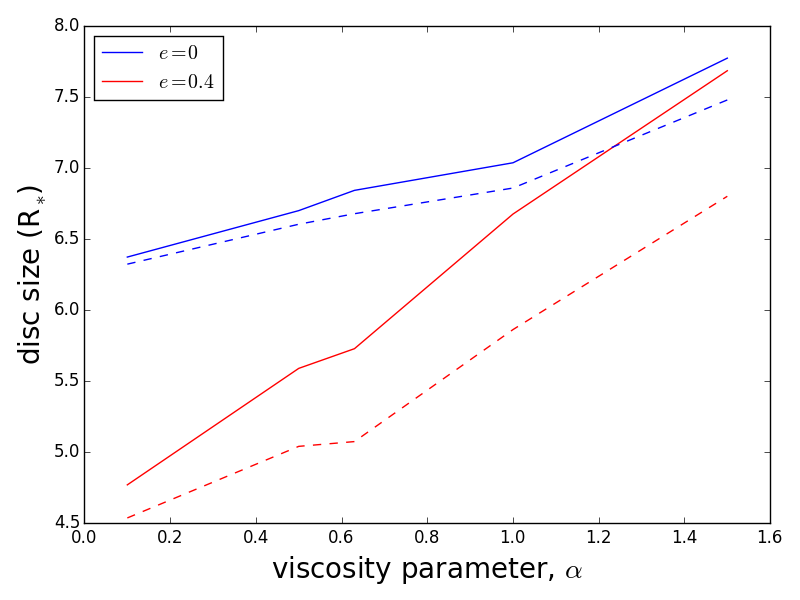}
	\caption{The relationship between the time-averaged size of the disc and viscosity parameter for systems with an orbital period of 40 days and eccentricities of $e=0$ and 0.4. The solid and dashed lines show the maximum and minimum disc sizes, respectively.}
	\label{fig:VISCsize}
\end{figure}

Figure \ref{fig:VISCsize} shows the variation of the Be star's disc size with viscosity. Minimum and maximum disc size both increase with viscosity for all eccentricities. Both the rate that disc size increases and the range from minimum to maximum are larger for higher eccentricities. For the viscosity parameters shown, the disc size increases by $\sim 60\%$.

\begin{figure}
	\centering
	\includegraphics[width=.5\textwidth]{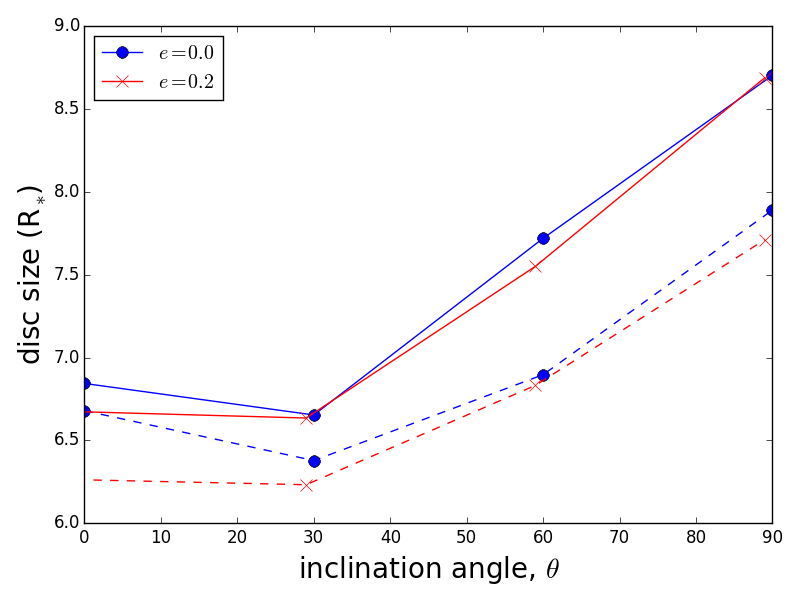}
	\includegraphics[width=.5\textwidth]{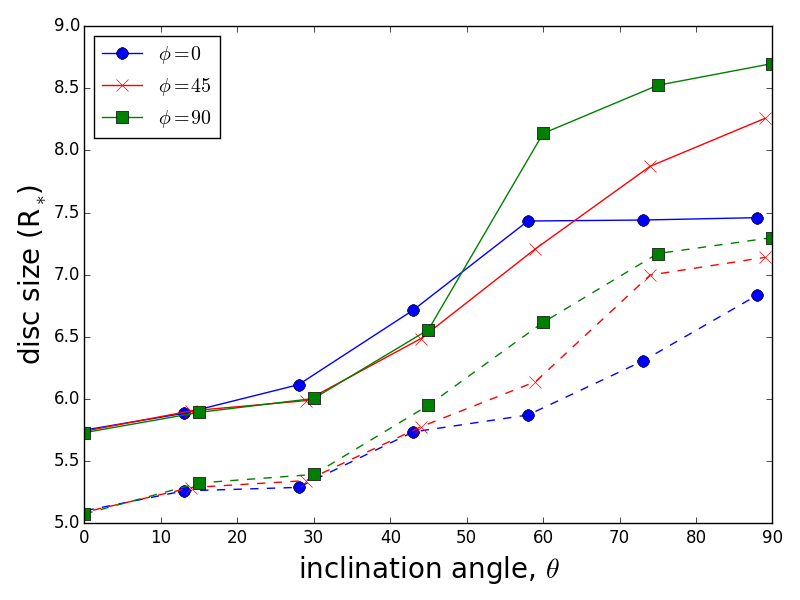}
	\caption{Top: Time-averaged size of the disc for various orientations. The systems shown have an orbital period of 40 days and eccentricities of $e=0.0$ and $0.2$. The solid and dashed lines show the maximum and minimum disc sizes, respectively. The values of maximum and minimum disc size are for simulations of any $\phi$ at each eccentricity. Bottom: Time-averaged size of the disc for various disc orientations. $\phi$ indicates the azimuthal rotation, i.e. rotation in the plane of the orbit. This is for systems with a 40 day period and 0.4 eccentricity. The solid and dashed lines show the maximum and minimum disc sizes, respectively.}
	\label{fig:INCLsize}
\end{figure}

Figure \ref{fig:INCLsize} shows the relationship between the disc's size and its orientation. Simulations were performed whilst varying both the inclination angle, $\theta$ and the azimuthal angle, $\phi$. Disc size increases with the inclination to the orbital plane and at $e=0.4$; the disc is $\sim40\%$ larger when the disc is perpendicular to the orbital plane than when it is coplanar. For $e = 0.0$ and $e=0.2$, the azimuthal angle has negligible effect on the size of the disc and hence the top panel of Figure \ref{fig:INCLsize} shows the minimum and maximum values of disc size for simulations of any azimuthal angle at each eccentricity. There is a relationship with azimuthal angle at higher eccentricities, as shown by the bottom panel of Figure \ref{fig:INCLsize}. The disc is at its largest at $\phi=90^{\circ}$, where the disc is perpendicular to the plane of the orbit but parallel to the semi-major axis. This is where the disc has minimum interaction with the neutron star, i.e. the neutron star passes close to the disc once at periastron. The disc is smaller when it is both perpendicular to the orbital plane and the semi-major axis ($\phi=0^{\circ}$). In this case, the neutron star interacts twice with the Be star's circumstellar disc: once before periastron and once after. Like viscosity, the range from minimum to maximum disc size increases with eccentricity. 

\begin{figure}
	\centering
	\includegraphics[width=.5\textwidth]{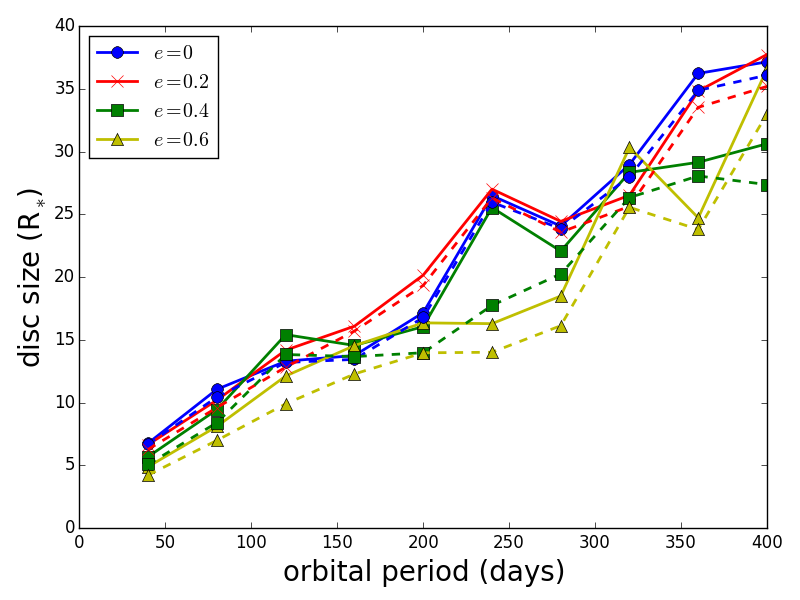}
	\caption{The relationship between the time-averaged size of the disc and the orbital parameters of the Be/neutron star binaries. The data points have a range of eccentricities from $e=0.0$ to $0.6$. The solid and dashed lines show the maximum and minimum disc sizes, respectively.}
	\label{fig:ORBPsize}
\end{figure}

Figure \ref{fig:ORBPsize} demonstrates the relationship between the orbital period of the Be/neutron star binary and the size of the Be star's circumstellar disc. The size of the disc increases by a factor of $\sim7$ with orbital period. The periastron distance of the neutron star and, in turn, the truncation radius of the Be star's disc increases with orbital period, allowing for a larger maximum disc radius. The difference between minimum and maximum disc size increases with eccentricity. 

Disc size is independent of the mass ejection rate of the Be star. A higher mass ejection rate increases the density of the disc but the radius of the tidal truncation remains the same and therefore so does the size of the disc. 

\section{Neutron star accretion rate} \label{sec:NSacc}

The X-ray emission from Be/neutron star binaries is a defining feature. The compact object companion is often only detectable by X-ray telescopes, and thus the X-ray flux is a vital quantity to understand in Be/neutron star binaries. In this paper, the neutron star mass capture rate is calculated directly from the number of simulation particles falling onto the neutron star. The accretion rate of the neutron star is assumed to be identical to the mass capture rate. The maximum accretion rate is shown because, unlike the average accretion rate, it is independent of the fraction of the orbit that the neutron star spends interacting with the Be star's circumstellar disc. Maximum accretion rate is calculated for individual orbits and then the median of five orbits is taken.


The neutron star's accretion rate increases linearly with the mass ejection rate of the Be star. A higher mass ejection rate yields a generally higher density of the disc (see Figure \ref{fig:INJEbgd}). When the truncation radius remains the same, the neutron star interacts at the same distance with a higher density disc. 

\begin{figure}
	\centering
	\includegraphics[width=.5\textwidth]{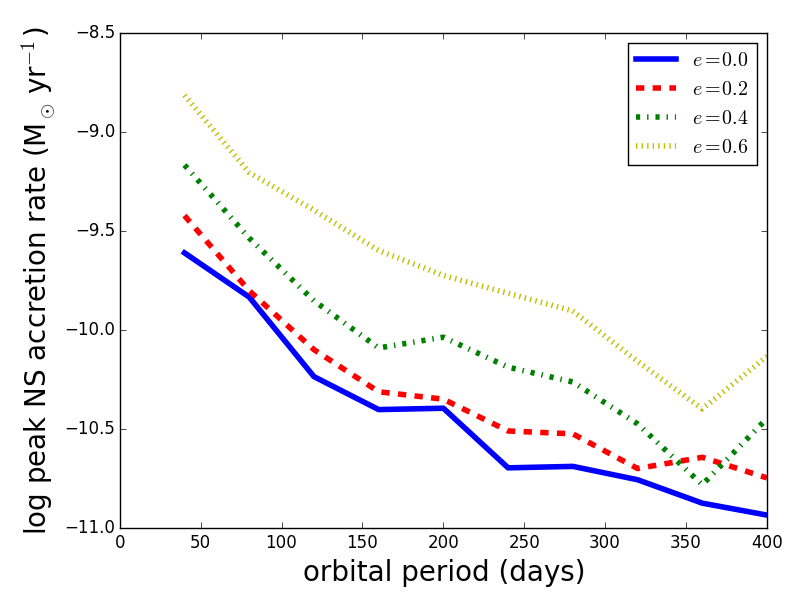}
	\caption{The relationship between the maximum accretion rate of the neutron star and the orbital period. }
	\label{fig:ORBPNSca}
\end{figure}

The relationship between the accretion rate of the neutron star and the orbital period of the Be/X-ray binaries with coplanar discs is shown in Figure \ref{fig:ORBPNSca}. The periastron distance of the neutron star is dependent on the orbital period and eccentricity of the binary. The density of the disc falls radially as a power law and thus the amount of accreted matter is dependent on the distance from the Be star. Therefore, neutron stars with larger orbital periods accrete less matter and the most eccentric systems have a higher maximum neutron star accretion rate.

The accretion rate of the neutron star has little to no dependency on the viscosity and the orientation of the disc. The orientation of the disc does not greatly alter the neutron star's closest passage to the disc and thus has a small effect on the maximum accretion rate. Larger viscosities both decrease the density of the disc (see Figure \ref{fig:VISCbgd}) and increase the size of the disc (see Figure \ref{fig:VISCsize}). This leads to a negligible change in the accretion rate of the neutron star.

\section{Comparison to observations} \label{sec:Obs}

CK15 presented two relationships involving the size of the Be star's circumstellar disc in Be/X-ray binaries (see Figures 8 and 10 in their paper). The first is a relationship between the H$\alpha$ equivalent width and the orbital period. They show two linear fits that utilise the sample of Be/neutron star binaries contained in the paper. One is a fit to the systems that possess an orbital period of $P_{\mathrm{orb}} \le 150$ days and the other is a fit to the entire observational sample. The former linear fit has a much larger gradient than the latter (see Table \ref{tab:FitsForDS}). The second relationship they describe is a quadratic relationship between the semi-major axis of the binary and the size of the disc. The disc size is determined using the following equation found in \citet{Hanuschik1989}:

\begin{equation} \label{eq:Hanuschik}
log \left( \sqrt{\frac{ R_{\mathrm{OB}} }{ R_{\mathrm{cs}} }}\right)  = [-0.32 \times log(-EW)] - 0.2 ,
\end{equation}

\noindent where $R_{\mathrm{cs}}$ is the size of the circumstellar disc, $EW$ is the H$\alpha$ equivalent width and $R_{\mathrm{OB}}$ is the radius of the Be star that is determined from the individual spectral types recorded for each Be star. In this section, these relationships are investigated using simulations with orbital periods ranging from 40 to 400 days and eccentricities of $e=0.0, 0.2, 0.4$ and $0.6$. The linear fits to the simulations, the observational data with $P_{\mathrm{orb}} \le 150$ days and the complete observational dataset will be referred to as $R_{\mathrm{sim}}$, $R_{150}$ and $R_{\mathrm{all}}$, respectively.

\begin{figure}
	\centering
	\includegraphics[width=.5\textwidth]{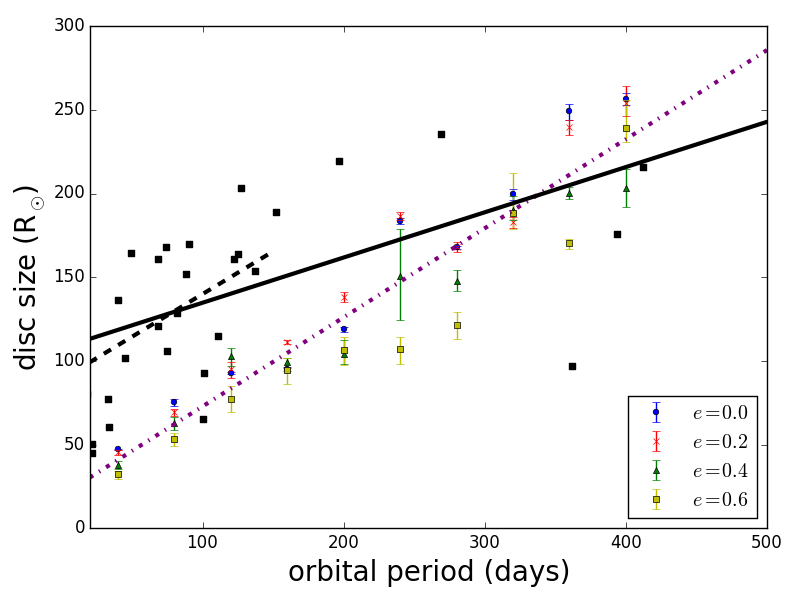}
	\caption{The relationship between disc size and orbital period. The black squares show the observational data from CK15. The coloured bars represent the simulation data and have a range of eccentricities from $e=0.0$ to $0.6$. The bars show the maximum and minimum values of disc size. The dot-dashed purple line shows the fit to all the simulation data. The fits from CK15 to systems with $P_{\mathrm{orb}} \le 150$ days (dashed black line) and systems with $P_{\mathrm{orb}} \le 500$ days (solid black line) are included in the plot. The parameters of the fits are described in Table \ref{tab:FitsForDS}.}
	\label{fig:ComparisonFig8}
\end{figure}

Figure \ref{fig:ComparisonFig8} shows the relationship between the size of the Be star's circumstellar and orbital period. The data and fits shown in Figure 8 of CK15 are included. The observational disc sizes are converted from the H$\alpha$ equivalent widths given in CK15 using Equation \ref{eq:Hanuschik}. Values for the Be star radii are taken from \citet{Schmidt1982}. The gradient of $R_{\mathrm{sim}}$ is equal to the gradient of $R_{150}$. However, the y-intercept differs by a factor of $\sim4$. Both the gradient and the y-intercept differ considerably between $R_{\mathrm{all}}$ and $R_{\mathrm{sim}}$. The discrepancy between the y-intercept of the simulations and the observations is most likely due to the method used to calculate the size of the disc. It is extremely difficult to determine the size of a Be star's disc from its emission because of the complex structure. Hence, the equation suggested by \citet{Hanuschik1989} is a relative value because it is determined by the emission of the disc. It is also developed from a sample of the brightest Be stars, largely including isolated Be stars. The discs in these systems are not truncated and hence have different density structures to those found in binary systems. This has an effect on the fraction of the disc that is emitting in a given wavelength and hence changes the relation between the emission and size of the Be star's disc.

\begin{table}
	\centering
	\caption{The gradients and y-intercepts for the linear fits shown in Figure \ref{fig:ComparisonFig8}. The three samples refer to the simulation data, the observational data with orbital periods of less than 150 days and the entire set of observational data. The data for the observed binaries is taken from CK15. }
    \label{tab:FitsForDS}
	\begin{tabular}{lcc} 
		\hline
		sample                                        & gradient & y-intercept (days)   \\
		\hline
		simulations                                   & 0.51     & 20.99          \\
		binaries with $P_{\mathrm{orb}}\leq$ 150 days & 0.51     & 88.90          \\
		complete observational sample                 & 0.27     & 107.70         \\
		\hline
	\end{tabular}
\end{table}

\begin{figure}
	\centering
	\includegraphics[width=.5\textwidth]{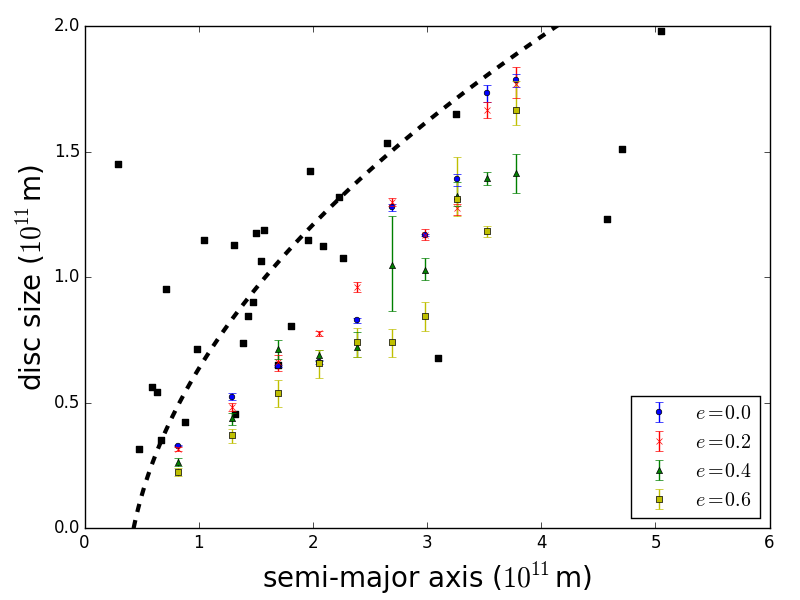}
	\caption{The relationship between the Be star's circumstellar disc size and semi-major axis of the neutron star's orbit. The black squares show the observational data from CK15. The coloured bars represent the simulation data and have a range of eccentricities from $e=0.0$ to $0.6$. The bars show the minimum and maximum values of disc size. The quadratic fit from CK15 (dashed black line) is given by Equation \ref{eq:quadFitObs}.}
	\label{fig:ComparisonFig10}
\end{figure}

Figure \ref{fig:ComparisonFig10} shows the relationship between the semi-major axis and the disc's size. For the simulations, disc size is calculated as described in Section \ref{sec:discSize} and semi-major axis is known from the assumed orbital period of each simulation. CK15 present the following quadratic fit

\begin{equation} \label{eq:quadFitObs}
a = (7 \times 10^{-12}) R_{\mathrm{cs}}^{2} + 0.4524 R_{\mathrm{cs}} + (4.3\times 10^{10}) \mathrm{m} .
\end{equation}

\noindent where $a$ is the semi-major axis and $R_{\mathrm{cs}}$ is the radius of the circumstellar disc in metres. The simulation data agrees with the observational relationship between the disc's size and the semi-major axis of the orbit. 

The simulation data in Figures \ref{fig:ComparisonFig8} and \ref{fig:ComparisonFig10} have an equal number of binaries at each tested eccentricity ($e = 0.0,0.2,0.4$ and $0.6$). The sample used by CK15 is missing a large number of the orbital eccentricities and hence there could be an unknown bias. This bias could be the cause of the differences between the observational and predicted relationships shown in this paper. 

In Figure \ref{fig:ComparisonFig8}, $R_{\mathrm{sim}}$ and $R_{150}$ have the same gradient. This implies that the observed binaries with low orbital periods possess a similar distribution of eccentricities to the simulation data. This is supported by the seven observationally determined eccentricities that are distributed over the range $0 \le e \le 0.5$ (see Table \ref{tab:knownEccs}). $R_{\mathrm{all}}$ has a considerably smaller gradient than $R_{\mathrm{sim}}$ and $R_{150}$. Removing all the simulations with larger orbital periods (i.e. $P_{\mathrm{orb}} > 150$ days) that have eccentricities of $e < 0.6$ lowers the gradient of the model fit to $0.41$. Hence, the simulations suggest that the eccentricities of the observational sample are higher at larger orbital periods. 

\begin{figure} 
	\centering
	\includegraphics[width=.5\textwidth]{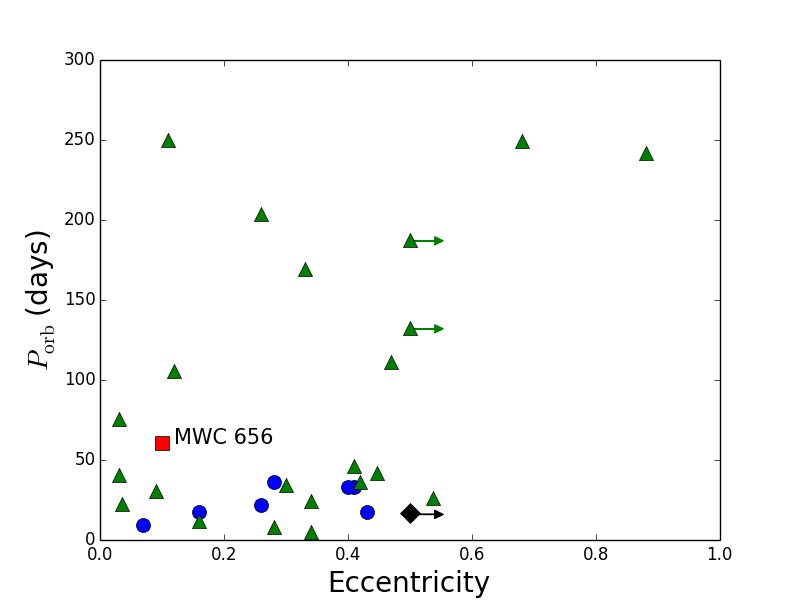}
	\caption{The relationship between eccentricity and orbital period for Be/X-ray binaries in the Milky Way, LMC and SMC. The only confirmed Be/black hole system, MWC 656, is also included. Arrows demonstrate the lower limit on eccentricity for the 3 systems with $e>0.5$. The values used for this figure are contained in \citet{BrownBH2018}.}
	\label{fig:EccVSPorb}
\end{figure}

Figure \ref{fig:EccVSPorb} shows that the distribution of eccentricities increases with orbital period for observed Be/X-ray binaries. Despite $80 \%$ of the systems having an orbital period of less than 150 days, the only two systems with a confirmed eccentricity greater than $0.6$ comprise two of the six binaries of larger orbital period. This supports the suggestion that the systems with higher orbital periods have a larger eccentricity on average. Evolutionary models suggest that, for a given mass loss and kick velocity during a supernova, $(a_{\mathrm{i}} / a_{\mathrm{f}}) \propto (1-e^{2})$ \citep{PostYung2014} where $a_{\mathrm{i}}$ and $a_{\mathrm{f}}$ represent the semi-major axis of the binary system before and after the supernova, respectively. Hence, the resultant orbital period of the binary is expected to increase with eccentricity, which agrees with the data shown in Figure \ref{fig:EccVSPorb}. A larger sample of observed Be/X-ray binaries with a high eccentricity is required to fully support this relationship.

Of the Be/X-ray binaries with known eccentricities, $\sim80 \%$ have an eccentricity of $e \le 0.45$ \citep{BrownBH2018}. CK15 present the eccentricity of seven systems in their sample and all of them are below $0.45$ (see Table \ref{tab:knownEccs}). It should be noted that these systems have orbital periods of $P_{\mathrm{orb}} < 40$ days. In Figure \ref{fig:ComparisonFig10}, the simulation data lies increasingly far away from the quadratic fit as the eccentricity grows. Therefore, it is suggested that the observational sample is more likely to contain binaries with low eccentricity.

\begin{table}
	\centering
	\caption{The orbital period and eccentricity of the seven systems with known eccentricities in the sample contained in CK15.}
    \label{tab:knownEccs}
	\begin{tabular}{lcc} 
		\hline
		system     & orbital period (days) & eccentricity    \\
		\hline
		SXP 2.37   & 9.30                  & 0.07 $\pm$ 0.02 \\
		SXP 5.05   & 17.20                 & 0.16 $\pm$ 0.02 \\
		SXP 6.85   & 22.00                 & 0.26 $\pm$ 0.03 \\
		SXP 8.80   & 33.40                 & 0.41 $\pm$ 0.04 \\
		SXP 11.5   & 36.30                 & 0.28 $\pm$ 0.03 \\
		SXP 18.3   & 17.79                 & 0.43 $\pm$ 0.03 \\
		SXP 74.7   & 33.30                 & 0.40 $\pm$ 0.23 \\
		\hline
	\end{tabular}
\end{table}

\section{Discussion and Conclusions} \label{sec:Conclusions}

In this paper, simulations are used to investigate three characteristics (the base gas density of the Be star's circumstellar disc, the accretion rate of the neutron star and the size of the disc) of Be/neutron binaries. Five parameters are varied (the viscosity and orientation of the Be star's circumstellar disc, the mass ejection rate of the Be star and the period and eccentricity of the neutron star's orbit) and the effect on the aforementioned characteristics is presented.

The base gas density is both dependent on the mass ejection rate of the Be star and the viscosity of the material in the disc. Using simulations, the density of the inner region of the disc has been shown to drop rapidly when the ejection rate of the Be star is decreased \citep{Haubois2012}. Our simulations also display this dependence of the base gas density on mass ejection rate. When the disc is in a steady state, the base gas density remains in equilibrium due to the balance between the ejection of matter into the disc and the accretion of matter back onto the Be star. Larger viscosities decrease the base gas density of the disc because viscous forces are diffusive. There is little to no dependence of the base gas density on the orientation of the disc or the orbital parameters of the binary. These findings agree with previous work that suggests that the orbital parameters of the binary only affect the truncation of the disc and not the inner regions \citep{OkazakiTrunc2002,Cyr2017}.

The disc's size in Be/X-ray binaries is limited by tidal truncation that arises from the presence of a binary companion. Hence, any properties that affect the interaction between the neutron star and the disc are important - this includes the orbital parameters and the orientation of the disc. Both orbital period and eccentricity vary the periastron distance of the neutron star's orbit and, in turn, increase (or decrease in the case of eccentricity) the size of the disc. The size of the disc increases with misalignment, in agreement with previous work that suggests a smaller tidal torque in binary systems with misaligned discs \citep{MartinBe2014,Cyr2017}. The disc's size increases with viscosity, in agreement with previous work that shows that the truncation radius is the distance where the tidal torque balances the viscous torque \citep{OkazakiTrunc2002,Panoglou2016}. The size of the Be star's disc is not dependent on the mass ejection rate of the Be star as both of the tidal and viscous torques are proportional to disc density (e.g., \citealt{ArtyLubow1991}; see also \citealt{NegOka2001}) and the disc is truncated at the same radius regardless of the density of the disc.

The maximum accretion rate of the neutron star is dependent on the mass ejection rate of the Be star, the orbital period and the orbital eccentricity. The density of the disc at the neutron star's closest passage to the Be star's disc determines the amount of material accreted. Given that the density in the disc falls off rapidly with increasing radius \citep{TouhGies2011}, the neutron star's maximum accretion rate is heavily dependent on the periastron distance of its orbit. 
Thus, the neutron star's maximum accretion rate is higher for systems with a smaller orbital period and higher eccentricity. The mass ejection rate of the Be star varies the overall density of the disc and thus it controls the density of the matter that the neutron star accretes. Thus, there is a linear relationship between the Be star's mass ejection rate and the neutron star's maximum accretion rate. The dependence of the maximum accretion rate on the orientation and viscosity of the disc is negligible.

The relationship between the size of the Be star's circumstellar disc, the orbital period and the semi-major axis were tested using simulations. The relationship between disc size and semi-major axis was similarly tested. In both cases, the simulation data agrees with the observational fits. The simulations suggest that the observational sample of Be/neutron star binaries with periods of $P_{\mathrm{orb}} \le 150$ days possess eccentricities that are distributed in the range $0.0 \le e \le 0.5$. It is also suggested that the binaries with larger orbital periods have a wider distribution of eccentricities and a larger average eccenctricity.


\section*{Acknowledgements}

ROB acknowledges support from the Engineering and Physical Sciences Research Council Centre for Doctoral Training grant EP/L015382/1. WCGH acknowledges support from the Science and Technology Facilities Council through grant number ST/R00045X/1. The authors acknowledge the use of the IRIDIS High Performance Computing Facility, and associated support services at the University of Southampton, in the completion of this work. Numerical computations were also performed on the Sciama High Performance Compute (HPC) cluster which is supported by the ICG, SEPNet and the University of Portsmouth. We thank the anonymous reviewer for their helpful comments.





\begin{thebibliography}{99}

\bibitem[\protect\citeauthoryear{Artymowicz and Lubow}{1991}]{ArtyLubow1991}
Artymowicz P. and Lubow S. H., 1991, \apj, 421, 651

\bibitem[\protect\citeauthoryear{Bate et al}{1995}]{Bate1995}
Bate M. R., Bonell I. A. and Price N. M., 1995, MNRAS, 277, 362-376
\bibitem[\protect\citeauthoryear{Benz}{1990}]{Benz1990}
Benz W., 1990, Numerical Modelling of Nonlinear Stellar Pulsations Problems and Prospects, 269 
\bibitem[\protect\citeauthoryear{Benz et al.}{1990}]{Benzetal1990}
Benz W., Bowers R. L., Cameron A. G. W. and Press, W. H., 1990, \apj, 348, 647
\bibitem[\protect\citeauthoryear{Brown et al.}{2018}]{BrownBH2018}
Brown R. O., Ho W. C. G., Coe M. J. and Okazaki A. T., 2018, MNRAS, 477, 4810
\bibitem[\protect\citeauthoryear{Brown et al.}{2019}]{Brown2019}
Brown R. O., Coe M. J., Ho W. C. G. and Okazaki A. T., 2019, MNRAS, 486, 3078

\bibitem[\protect\citeauthoryear{Carciofi and Bjorkman}{2006}]{CarcBjor2006}
Carciofi A. C. and Bjorkman J. E., 2006, \apj, 639, 1081
\bibitem[\protect\citeauthoryear{Carciofi and Bjorkman}{2008}]{CarcBjor2008}
Carciofi A. C. and Bjorkman J. E., 2008, \apj, 684, 1374
\bibitem[\protect\citeauthoryear{Coe and Kirk}{2015}]{CoeKirk2015}
Coe M. J. and Kirk J., 2015, MNRAS, 452, 969
\bibitem[\protect\citeauthoryear{Coe et al.}{2015}]{Coe2015}
Coe M. J. et al., 2015, MNRAS, 447, 2387
\bibitem[\protect\citeauthoryear{Coleiro et al.}{2013}]{Coleiro2013}
Coleiro A. et al., 2013, A\&A, 560, 108
\bibitem[\protect\citeauthoryear{Cox}{2000}]{Cox2000}
Cox A. N., 2000, Allen's Astrophysical Quantities (New York: Springer-Verlag)
\bibitem[\protect\citeauthoryear{Cyr et al.}{2017}]{Cyr2017}
Cyr I. H., Jones C. E., Panoglou D., Carciofi A. C. and Okazaki A. T., 2017, MNRAS, 471, 596

\bibitem[\protect\citeauthoryear{Dougherty and Taylor}{1992}]{DouTay1992}
Dougherty S. M. and Taylor A. R., 1992, \nat, 359, 808 


\bibitem[\protect\citeauthoryear{Ghoreyshi et al.}{2018}]{Ghor2018}
Ghoreyshi M. R. et al., 2018, MNRAS, 479, 2214

\bibitem[\protect\citeauthoryear{Hanuschik}{1989}]{Hanuschik1989}
Hanuschik R. W., 1989, APSS, 161, 61
\bibitem[\protect\citeauthoryear{Haubois et al.}{2012}]{Haubois2012}
Haubois X., Carciofi A. C., Rivinius T., Okazaki A. T., Bjorkman J. E., 2012, Astrophys. J., 756, 156

\bibitem[\protect\citeauthoryear{Jaschek and Egret}{1982}]{JascEgre1982}
Jaschek M. and Egret D., 1982, Proceedings IAU Symposium No. 98, Dordrecht: D. Reidel Publishing Co., 261

\bibitem[\protect\citeauthoryear{Kiseleva et al.}{1998}]{Kiseleva1998}
Kiseleva L. G., Eggleton P. P. and Mikkola S. 1998, MNRAS, 300, 292
\bibitem[\protect\citeauthoryear{Kozai}{1962}]{Kozai1962}
Kozai Y., 1962, AJ, 67, 591
\bibitem[\protect\citeauthoryear{K{\"u}hnel et al.}{2017}]{Kuhnel2017}
K{\"u}hnel M. et al., 2017, MNRAS, 471, 1553

\bibitem[\protect\citeauthoryear{Lidov}{1962}]{Lidov1962}
Lidov M. L., 1962, Planet. Space Sci., 9, 719

\bibitem[\protect\citeauthoryear{Martin et al.}{2014}]{MartinBe2014} 
Martin R. G., Nixon C., Armitage P. J., Lubow S. H. and Price D. J., 2014, ApJL, 790, 34

\bibitem[\protect\citeauthoryear{Negueruela and Okazaki}{2001}]{NegOka2001}
Negueruela I. and Okazaki A. T., 2001, A\&A, 369, 108

\bibitem[\protect\citeauthoryear{Okazaki et al.}{2002}]{OkazakiTrunc2002}
Okazaki A. T., Bate M. R., Ogilvie G.I. and Pringle J. E., 2002, \pasp, 261, 519 

\bibitem[\protect\citeauthoryear{Panoglou et al.}{2016}]{Panoglou2016}
Panoglou D. et al.\ 2016, \mnras, 461, 2616 
\bibitem[\protect\citeauthoryear{Poeckert and Marlborough}{1982}]{PoeMarl1982}
Poeckert R. and Marlborough J. M., 1982,  Astrophys. J., 252, 196
\bibitem[\protect\citeauthoryear{Postnov and Yungelson}{2014}]{PostYung2014}
Postnov K. A. and Yungelson L. R., 2014, LRR, 17, 3
\bibitem[\protect\citeauthoryear{Postnov et al.}{2014}]{Postnov2014}
Postnov K. A. et al., 2014, MNRAS, 446, 1013


\bibitem[\protect\citeauthoryear{Rappaport and van de Heuvel}{1982}]{RapHeu1982}
Rappaport S. and van de Heuvel E. P. J., 1982, IAU Symposium  No. 98 \textit{Be Stars} (M. Jaschek and H.G. Groth, Editors), Reidel, Dordrecht, p. 327
\bibitem[\protect\citeauthoryear{Reig et al.}{2011}]{Reig2011}
Reig, P.\ 2011, \apss, 332, 1
\bibitem[\protect\citeauthoryear{Reig, Fabregat and Coe}{1997}]{ReigFabCoe1997}
Reig P., Fabregat J., Coe M. J., 1997, A\&A, 322, 193
\bibitem[\protect\citeauthoryear{R\'{i}mulo et al.}{2018}]{Rimulo2018}
R\'{i}mulo L. R. et al., 2018, MNRAS, 476, 3555
\bibitem[\protect\citeauthoryear{Rivinius}{2000}]{Riv2000}
Rivinius T., 2000, ASP Conference Series, 214, 228
\bibitem[\protect\citeauthoryear{Rivinius, Carciofi and Martayan}{2013}]{RivCarc2013}
Rivinius T., Carciofi A. C. and Martayan C., 2013, A\&A Review, 21, 69
\bibitem[\protect\citeauthoryear{Ru\v{z}djak et al.}{2009}]{Ruzdjak2009}
Ru\v{z}djak D. et al., 2009, A\&A, 506, 1319

\bibitem[\protect\citeauthoryear{Schmidt-Kaler}{1982}]{Schmidt1982}
Schmidt-Kaler Th., 1982, in Schaifers K., Voigt H. H., eds, Landolt-B\:{o}rnstein, group VI, Vol 2, subvol. b, Stars and Star Clusters. Springer, Berlin, Heidelberg, New York, p. 17
\bibitem[\protect\citeauthoryear{Shakura and Sunyaev}{1973}]{ShakSuny1973}
Shakura N. I. and Sunyaev R. A., 1973, A\&A, 24, 337
\bibitem[\protect\citeauthoryear{Slettebak}{1988}]{Slettebak1988}
Slettebak A., 1988, Publ. Astron. Soc. Pac, 100, 770

\bibitem[\protect\citeauthoryear{Touhami, Gies and Schaefer}{2011}]{TouhGies2011}
Touhami Y., Gies D. R. and Schaefer G. H., 2011, Astrophys. J., 729, 17


\bibitem[\protect\citeauthoryear{van de Heuvel and Rappaport}{1987}]{HeuRap1987}
van den Heuvel E.P.J. and Rappaport S., 1987, IAU Colloquium No. 92 \textit{Physics of Be Stars} (A. Slettebak and T.P. Snow, editors), Cambridge University Press, p. 291
\bibitem[\protect\citeauthoryear{van Kerkwijk et al.}{1995}]{vanKer1995}
van Kerkwijk M. H., Waters L. B. F. M. and Marlborough J. M., 1995, A\&A, 300, 259
\bibitem[\protect\citeauthoryear{Vieira et al.}{2015}]{Vieira2015}
Vieira R. G. et al., 2015, MNRAS, 454, 2107

\end{thebibliography}








\label{lastpage}
\end{document}